\documentclass[12pt,preprint]{aastex}

\slugcomment{}

\shorttitle{}
\shortauthors{}

\begin{document}


\title{The Failure of Self-Interacting Dark Matter to solve the Overabundance of 
Dark Satellites and the Soft Core Question}


\author{Elena D'Onghia\altaffilmark{1} and Andreas Burkert\altaffilmark{1}}
\affil{Max-Planck-Institut f\"ur Astronomie, K\"onigstuhl 17, 69117 Heidelberg,
Germany}






\begin{abstract}
Self-interacting dark matter  was proposed by Spergel $\&$ Steinhardt
(2000) to alleviate two conflicts between  Cold Dark Matter (CDM)
models and observations. Firstly,  CDM N-body simulations predict
dark matter halo density profiles that diverge at the centre 
in disagreement with the constant 
density cores observed in late-type dwarf and Low Surface
Brightness (LSB) galaxies. Secondly,  N-body simulations
predict an overabundance of subhalos in the Galactic halo.
Using a simple semi-analytical argument we show that 
 weakly self-interacting dark matter models, which can  
produce halo cores of the sizes observed in dark 
matter dominated galaxies, are unable to reconcile the number of satellites 
in the Galactic halo with the observed number of dwarf galaxies 
in the Local Group. 
\end{abstract}


\section{Introduction}
Recent improvements in observational and numerical techniques have 
allowed a comparison between predictions of the CDM scenario and 
observational data on galactic scales. The results point out 
discrepancies between predictions and observations. 
High-resolution N-body simulations have shown that, on scales 
comparable to the Local Group, the predicted number of subhalos 
is at least a factor of ten higher than the observed 
number of dwarf galaxies (Klypin et al. 1999, Moore et al. 1999a).
This disagreement, usually called the ``satellite question'', can be 
attributed to the high core densities of satellite dark halos
found in cosmological models (Navarro, Frenk $\&$ White 1997; hereafter NFW). 
These densities, combined with a small central velocity
dispersion (Fukushige $\&$ Makino 1997), tend to stabilize
the satellites against tidal disruption on galactic scales.
Another discrepancy emerges when comparing the density
profiles of dark matter halos predicted by numerical
simulations with observations of HI rotation curves
in dwarf galaxies (Moore 1994; Flores $\&$ Primack 1994;
Burkert 1995). Whereas observations show linearly
rising rotation curves out to radii greater 
than 1 $h^{-1}$ kpc, indicating that the dark matter has a 
constant density core (soft core), cosmological simulations predict 
dark halo 
density profiles with $\rho \propto r^{-1.5}$ in the central parts 
(Moore et al. 1999b; Fukushige $\&$ Makino 2001). Other N-body simulations 
appear to converge to halo density profiles described by  
$\rho \propto r^{-1}$ (Power et al. 2002). 
These two conflicts, which might be related, the excess of dark satellites and the soft core
question, arise because the CDM N-body simulations predict 
dark matter halos with high core densities.

Each conflict taken individually may not be sufficient
to invalidate CDM on galactic scales.
Results derived from observed density profiles of the inner regions
in galaxies are controversial,
due to beam smearing effects in HI rotation curves
(van den Bosch $\&$ Swaters 2001),
even though high-resolution observations of H$\alpha$ also show
shallower core densities than those predicted by CDM
numerical simulations (e.g. de Blok $\&$ McGaugh $\&$ Rubin 2001; Marchesini
et al. 2002).
Several authors have attempted to reconcile
the number of observed Local Group dwarf galaxies
with the number predicted by CDM theory through conservative solutions
within the framework of the current theory. 
Early work by Kauffmann, White $\&$ Guiderdoni (1993), using semi-analytic models
of galaxy formation, found most subhalos lacked a luminous component.
Energetic mechanisms which are more
efficient in low mass systems, such as feedback from evolving stars
and heating by an ionizing UV background, were proposed to explain
a decoupling of luminous and dark components for low mass dwarfs
(Efstathiou 1992, Bullock et al. 2000, Gelato $\&$Sommer-Larsen 1999, 
Thacker $\&$ Couchman 2000). 
Another solution, proposed by
Klypin (1999), suggested an identification of the missing satellites
seen in numerical simulations with observed compact
high-velocity clouds (Blitz et al. 1999). This proposal may
be premature, since it is still unclear  whether
the high-velocity clouds
are galactic or extragalactic in nature. Comparisons of the dark satellite
halos in CDM dominated simulations to the distribution of observed neutral
hydrogen high-velocity clouds and compact high-velocity clouds were 
made by Putman $\&$ Moore (2001). 
Recently, Stoehr et al. (2002) and Hayashi et al. (2002)
 suggested
 that the Galactic satellites could be identified with the most 
massive subhalos of CDM simulations. This would tend to support 
scenarios in which baryons are lost preferentially from low-mass halos for yet
 unknown reasons.

Still, the disagreement between observations and predictions might indicate
that a revision to the CDM scenario is required.
Self-interacting dark matter was proposed by Spergel $\&$ Steinhardt
(2000) to overcome the satellite question and the soft core question.
In this model, dark matter particles experience  weak, non-dissipative,
collisions on scales of kpc to Mpc for typical galactic densities.
These collisions thermalize the inner regions of the
dark halos, producing a soft core. In addition, the excess of subhalos
predicted by the CDM models would be reduced.
This model has attracted great attention. Numerical simulations (e.g. Burkert 2000;
Yoshida et al. 2000; Moore et al. 2000; 
Firmani, D'Onghia $\&$ Chincarini 2001;  Dav\'{e} et al. 2001; D'Onghia, Firmani $\&$ Chincarini 2002)
demonstrated that, in this scenario, soft cores would form naturally after a collisional timescale 
by energy transport
into the cold inner regions. However, after the initial expansion, 
 the cores of isolated halos would evolve towards the core-collapse stage, with
 final central densities higher than those predicted by NFW 
(Burkert 2000; Kochanek $\&$ White 2000).
Ostriker (2000) and Hennawi $\&$ Ostriker (2002)
pointed out that self-interacting dark matter
in a very weak cross section
regime in the centers of galaxies reproduces supermassive
black hole masses
and their observed correlation
with the velocity dispersion of the host bulges. However, they point out a possible
inconsistency of the collisional scenario; indeed, the model would lead to the exorbitant
growth of supermassive black holes, which consequently  imposes a very strict upper limit on
the collisional cross section. Other limits on the cross section values
were derived from the fundamental plane relation for ellipticals in clusters
(Gnedin $\&$ Ostriker 2001). 

Using an analytical approach, this Letter explores whether 
weakly self-interacting dark matter is likely to reconcile the apparent
overabundance of subhalos with the small number of visible satellites 
in the Local Group. Two different disrupting processes are explored: collisions
and tidal stripping.  
This work assumes $H_0=70$ km s$^{-1}$ Mpc$^{-1}$. 

\section{Disruption by Collisions}
For a satellite dwarf galaxy orbiting the Milky Way, 
we define $\tau$ to be the time for the dark satellite halo to be 
destroyed by collisions with the 
self-interacting dark matter within the halo of the Milky Way:
\begin{equation}
\tau = \frac{1}{\rho_{MW} \ \sigma \ v}
\end{equation}
where $\rho_{MW}$ is the Galactic dark halo density (local density),
$\sigma \equiv \sigma_{si}/m_x$ is the self-interacting cross section per unit mass
and $v$ is the velocity of the satellite relative to the Milky Way.
We identify $v$ as the typical Milky Way halo velocity dispersion and 
assume that one collision for each particle of the satellite 
is enough to disrupt the satellite within a Hubble time. This assumption
implies an optically-thin regime (Gnedin $\&$ Ostriker 2001).

Let us assume a cross section inversely proportional
to the halo velocity dispersion. This choice for the cross 
section  produces smaller, less spherical cores in clusters
of galaxies and large cores in dwarf galaxies (Yoshida et al. 2000, 
Firmani et al. 2001, Wyithe, Turner $\&$ Spergel 2001),  consistent
with  observations of cluster cores like Cl 0024+1654. 
It also implies that
the product of the cross section times the halo dispersion velocity
is constant and independent of the mass. 
Thus, in the satellite the cross section
$\sigma'$ times the satellite dispersion velocity $v_0$ 
has the same value as the product $\sigma \cdot v$ in the Milky Way.

Let us suppose that the self-interacting dark matter of the satellite is interacting
with itself and at the same time with the Milky Way halo. 
In the satellite, the effect will be a halo central density decreasing due
to the collisions between dark matter particles.
Within a Hubble time $t_H$, the expected average collision time is:

\begin{equation}
\frac{t_H}{N_{\rm{coll}}}=\frac{1}{\rho_0 \ \sigma \ v},
\end{equation}

\noindent
Substituting eq.(2) in (1), under the hypothesis that
$\sigma \cdot v$ is constant, $\tau$ is given by:
\begin{equation}
\tau=\Big(\frac{\rho_0}{\rho_{MW}}\Big) \frac{t_H}{N_{\rm{coll}}}.
\end{equation}
We assume for the satellite halo central density the same value
observed in late-type dwarf galaxies: $\rho_0 \approx 0.02 \ M_{\odot}$pc$^{-3}$
(de Blok, McGaugh $\&$ Rubin 2001, Marchesini et al. 2002),
since the choice of $\sigma \propto 1/v$ predicts halo
central densities independent of the mass. 
The average collision rate (the inverse of eq.(2))
is a function of the halo central density,
the cross section and the halo velocity dispersion.
Since in this case  
$\sigma \cdot v$ is constant, the average collision rate is only a function 
only of the satellite central density: $N_{\rm{coll}}/t_H \propto \rho_0$.
Cosmological N-body simulations in which
the cross section is assumed to be inversely proportional 
to the halo velocity dispersion estimate $N_{\rm{coll}}\approx 3-4$
for each particle in the core 
for $\sigma \cdot v \approx 0.6$ cm$^2$ 
g$^{-1}$ in order to reproduce the 
central densities observed in late-type dwarf galaxies over a Hubble time
(D'Onghia, Firmani$\&$ Chincarini 2002).
The same estimates are obtained using a dynamical code
based on the integration of the Boltzmann equation 
 for the same value of the scattering cross section
(Firmani, D'Onghia $\&$ Chincarini 2001).

Let us consider a dwarf halo placed at a distance of 25 $h^{-1}$ kpc
from the centre of the Galactic halo. At this radius
the Milky Way density is predicted to be 
$\rho_{MW} \approx 4 \cdot 10^{-3}$ M$_{\odot}$pc$^{-3}$ (Moore et al. 2001).
Hence $\rho_0$ is nearly 5 times larger than the Galactic halo density 
$\rho_{MW}$. For $N_{\rm{coll}}=4$ the satellite disruption time 
at 25 $h^{-1}$ kpc is $\tau\approx t_H$. However, at only  
30  $h^{-1}$ kpc from the centre,  $\rho_0$ is 10 times larger than
the Milky Way density, producing 
$\tau\approx 2 t_H$$\footnote{In eq.(3) the density profile of the Milky Way, $\rho_{MW}$, 
in CDM models was adopted from Moore et al. 2001, Fig.2.
In that work, Moore and collaborators (2001) show
that the Milky Way dark density profile in CDM models is well described by:
$\rho(r) \propto 1/[(r/r_s)^{1.5}+(r/r_s)^3]$ with $r_s$  the scale length.}$.
In Figure 1 the time required to destroy satellites in the Galactic
halo is shown as a function of the distance from the centre
of the Milky Way as predicted by our analytic consideration when
$\sigma \propto 1/v$ is assumed (filled circles). Note that,  
at a distance of  50 $h^{-1}$ kpc, 10 Hubble times 
are required to destroy the satellites, if self-interaction is 
working to produce the central density we observe in late-type dwarf
galaxies. 

Let us now analyse the case in which the cross section is independent
of velocity: $\sigma \approx$ constant.
This case has interesting implications for 
supermassive black hole formation (Ostriker 2000). 
Eq.(3) becomes:
\begin{equation}
\tau=\Big(\frac{\rho_0}{\rho_{MW}}\Big)
     \Big(\frac{t_H}{N_{\rm{coll}}}\Big) 
     \Big(\frac{v_0}{v}\Big),
\end{equation}
with $v_0$ the satellite velocity dispersion.
For $\sigma \approx$constant  the halo central densities are no longer independent
of the halo mass and $\rho_0 v_0=\rho_0^{MW} v$
with $\rho_0^{MW}$ the central density of the Milky Way.
Since  the halo central density   
times the halo dispersion velocity is constant, the average collision rate 
is a function only of the cross section: $N_{\rm{coll}}/t_H \propto \sigma$.
Cosmological N-body simulations with velocity-independent cross sections 
of  $\sigma \sim 0.6$ cm$^2$ g$^{-1}$ 
suggest a central density of $\rho_0 \approx 4 \cdot 10^{-3}\ M_{\odot}$pc$^{-3}$
for satellites of $M=9\cdot 10^8 \ M_{\odot}$ (Dav\`{e} et al. 2001), 
and a typical number of collisions for each 
particle in the halo core of $N_{\rm{coll}} \simeq 3-4$ for $\sigma=0.1$ cm$^{2}$ g$^{-1}$
(Yoshida et al. 2000).
Thus, in eq.(4)  
we assume $N_{\rm{coll}}=4$ and $\rho_0=4 \cdot 10^{-3}\ M_{\odot}$pc$^{-3}$.
In Figure 1 the disruption time of satellites at different radii from the 
centre of the Milky Way is shown for $\sigma \approx$constant  
(open circles). For dwarf galaxies placed at a distance larger than 100 $h^{-1}$kpc, 
collisions between dark particles are inefficient in destroying 
satellites.

Satellite orbits are in general eccentric.
As a result, subhalos can be destroyed efficiently when 
their pericentric distances  are within 25 $h^{-1}$ or 100 $h^{-1}$kpc, 
depending on whether $\sigma$ is proportional to $1/v$ or not.
Let us concentrate on the case in which the cross section decreases with the
halo velocity dispersion, since it is in better agreement with the observed
size of soft cores in dwarf galaxies.
Using a Monte Carlo method we have computed the pericentric distance distribution
of the dark satellites found by CDM N-body simulations, assuming orbits with the same
eccentricity distribution function found for the halo orbits of cosmological N-body  simulations 
(Ghigna et al. 1998). These simulations yield very eccentric satellite orbits 
with pericentric over apocentric distance ratios of 
$R_{peri}/R_{apo}$$\sim$0.2, 
whereas observational evidence indicates that dwarf satellite orbits in the Local Group 
are more circular: $R_{peri}/R_{apo}=0.5$ (Schweitzer et al. 1995). 
Assuming the eccentric orbits, the chances are high for 
dark satellites to
move through the inner regions with pericentric distances $\le$ 25 $h^{-1}$ kpc 
and to be destroyed. 
Since satellites spend most of their
time at their apocentric distances, we assume that the semi-major axes of their
orbits are distributed in the same way satellite distances from the Milky Way centre are
predicted by N-body simulations.  
For each satellite predicted by CDM models, knowing its 
galactocentric  distance (courtesy B.Moore), 
 we compute with the Monte Carlo technique: (i) its probability of having an eccentricity $e$
predicted by the distribution function found  for the halo orbits of  N-body simulations
by Ghigna et al. (1998); (ii) its probability of having a semi-major axis $a$ as predicted by 
CDM models and from that its pericentric distance, $R_{peri}=a(1-e)$. All the satellites
with pericentric distances less then 25 $h^{-1}$kpc are assumed to be destroyed
by collisions.

In order to check the validity of our
model we have compared the final
pericentric distance distribution resulting from our Monte Carlo realization to that 
found in CDM models by Font and co-workers (2001). 
In Figure 2 the masses and pericentric distances of the subhalos  
within twice the virial radius are shown, resulting from our Monte Carlo realization.
The good agreement with the same plot shown in Font et al. (2001) is very
encouraging. 

What is the percentage of disrupted satellites
when the orbital eccentricity distribution is taken into account?
In Figure 3 the dashed
line shows the cumulative number of dwarf galaxies observed 
within 250 $h^{-1}$kpc from the Milky Way centre 
(Grebel 2000)$\footnote{The original distances of dwarf galaxies 
from the Sun in Grebel (2000) are corrected here for galactocentric distances.}$. 
The solid line indicates the cumulative
number of satellites predicted by SCDM models at the same radii$\footnote{The number of subhalos
predicted by Standard CDM models is from B. Moore. High-resolution N-body simulations
of $\Lambda$CDM models predict the same excess of dark satellites in the Local Group
(B. Moore, private communication).}$. 
Note that the N-body simulations predict an excess of subhalos at all 
radii. The dotted line represents the cumulative number of
substructures which should survive collisions, because 
their pericentric distances are larger than 25 $h^{-1}$ kpc.
We still find too many satellites beyond of 25 $h^{-1}$ kpc.
It is clear from Figure 3 that, for $\sigma \approx$1/v, collisional processes are not 
efficient enough in destroying substructures at any radii. 

If the 
cross section is assumed to be independent of the relative velocity
of the particles, then satellites with pericentric distances within 100 $h^{-1}$ kpc
will be destroyed by collisions.
Hence, no satellites should exist at radii smaller than
100 $h^{-1}$ kpc, in conflict with observations, while the overabundance of satellites
at larger radii remains insolved (triangles in Figure 3). Note that we neglect those satellites that
 spend  a very  short time within 100 $h^{-1}$ kpc; the likelihood of our detecting these 
satellites at such pericentric distances is small. 

\section{Disruption by Tidal Stripping}
In a self-interacting scenario
the halo central densities are expected to be lower than in CDM models.
As a result, tidal stripping should be more rapid and efficient and 
substructure halos
orbiting in the tidal field of the Milky Way are expected to continuously lose mass
and to be destroyed as a result of tidal forces. Thus, tidal stripping could 
be the dominant mechanism for destroying subhalos in a weakly self-interacting scenario.
We take this process into account by assuming that the orbiting satellite is tidally
truncated at some radius $r_t$, where the differential tidal force
of the Milky Way is equal to the gravitational attraction of the satellite.
For non-circular orbits, an espression  for the tidal radius 
may be derived from the discussion in Spitzer (1967, page 105):\\
\begin{equation}
\frac{m(r_t)}{r_t^3}=\Big(3+e \Big)\frac{M(R_{peri})}{R_{peri}^3} 
\end{equation} 
where $R_{peri}$ is the satellite pericentric distance, $m(r_t)$ the substructure mass 
within the tidal radius, $M$ the host galaxy mass and $e$ the satellites
orbital eccentricity. 
Thus the tidal radius is such that the mean density of
 the satellite within $r_t$ is proportional to the mean density of
the main halo at pericentric distance.
In  CDM satellites, when the tidally imposed radius approaches a value 
smaller than the scale radius $r_s$, substructures become unstable. 
Using N-body simulations of tidal stripping, the evolution of substructure halos 
described by a Hernquist or NFW profile 
within a static host potential have been explored by Mayer et al. (2001) and Hayashi
et al. (2002).

In the  
self-interacting scenario, both dark satellites and the Milky Way have lower
central densities than their CDM counterparts, and have core radii $r_c$.
 Substructures can be tidally stripped
when  their tidally imposed radii approach values  
smaller than their core radii $r_c$. 
 A detailed study of this mechanism requires numerical N-body simulations
and is beyond the scope of this work. However, we note that,
choosing a self-interacting cross section  of $\approx$0.6 cm$^2$ g$^{-1}$ 
which is capable of reproducing  the central density of 0.02 $M_{\odot}$pc$^{-3}$ 
and $r_c \approx 2 \ h^{-1}$ kpc for dark halos
 observed in late-type dwarf galaxies,  
 eq.(5) is satisfied
for substructures with pericentric distances $R_{peri}\le 40 \ h^{-1}$ kpc.
Following the same Monte Carlo procedure described 
in the previous section to derive the pericentric distance distribution for 
satellite orbits, we have determined  
the percentage of dark satellites found in the  SCDM simulations
 with $R_{peri}\le 40 \ h^{-1}$ kpc. The satellites with 
pericentric distances 
less than  40 $h^{-1}$ kpc
are assumed to be disrupted by tidal forces.

Filled points in Figure 3
show the cumulative number of substructures surviving the disruption 
by tidal stripping. Note that in a weakly self-interacting scenario capable of solving
the soft core question, tidal stripping  
is more efficient than collisions  in destroying subhalos with pericentric distances 
within 40 $h^{-1}$ kpc and thus alleviating the excess of satellites at small scales. 
However, tidal forces cannot solve the excess satellite problem at larger radii.

\section{CONCLUSION}

In a hierarchical universe, high-resolution N-body simulations
of Standard CDM models predict
an excess of subhalos with respect to the number of dwarf galaxies
observed in the Local Group (Klypin et al. 1999, Moore et al. 1999a). 

To solve this conflict between predictions and observations
a successful theory should reduce  the abundance
of substructures at all radii.
However, if the `satellite question' remains a problem
of CDM models, than 
our semi-analytical argument proves that self-interacting
dark matter is unable to solve it. 

If the value of the cross section
is chosen such that it reproduces
soft cores with sizes observed in late-type dwarf
and LSB galaxies and  assumed to decrease with the
halo velocity dispersion,
then collisions between particles
are  effective in disrupting subhalos only within  25 $h^{-1}$ kpc 
from the Milky Way centre. As a result, only a small percentage 
of all substructures are destroyed by collisions and the overabundance is only slightly reduced,
leaving the problem unsolved at all radii.
Tidal stripping is more efficient than collisions in
destroying subhalos within 40 $h^{-1}$ kpc from the Galactic centre, alleviating the problem
at small radii. However, discrepancy between predictions and 
observations persists. 

In summary, finding a process that is able to decrease the halo central density
does not appear  to be sufficient for reducing the excess of dark satellite halos, especially
for substructures placed at large distances from the Milky Way centre.  
Thus, weak self-interaction, which was originally
proposed to solve the soft core question in centres of  dark matter
dominated galaxies and the overabundance of subhalos in the Local
Groups is unable to solve both questions simultaneously. 

\acknowledgments

We are especially indebted to Ben Moore for providing the table with
the number of satellites predicted within the Milky Way halo in SCDM models
and their distances from Galactic centre.
Many thanks to Eva Grebel for useful discussions and to the referee Oleg Gnedin
for his suggestions and helpful comments.

\clearpage


\begin{figure}
\plotone{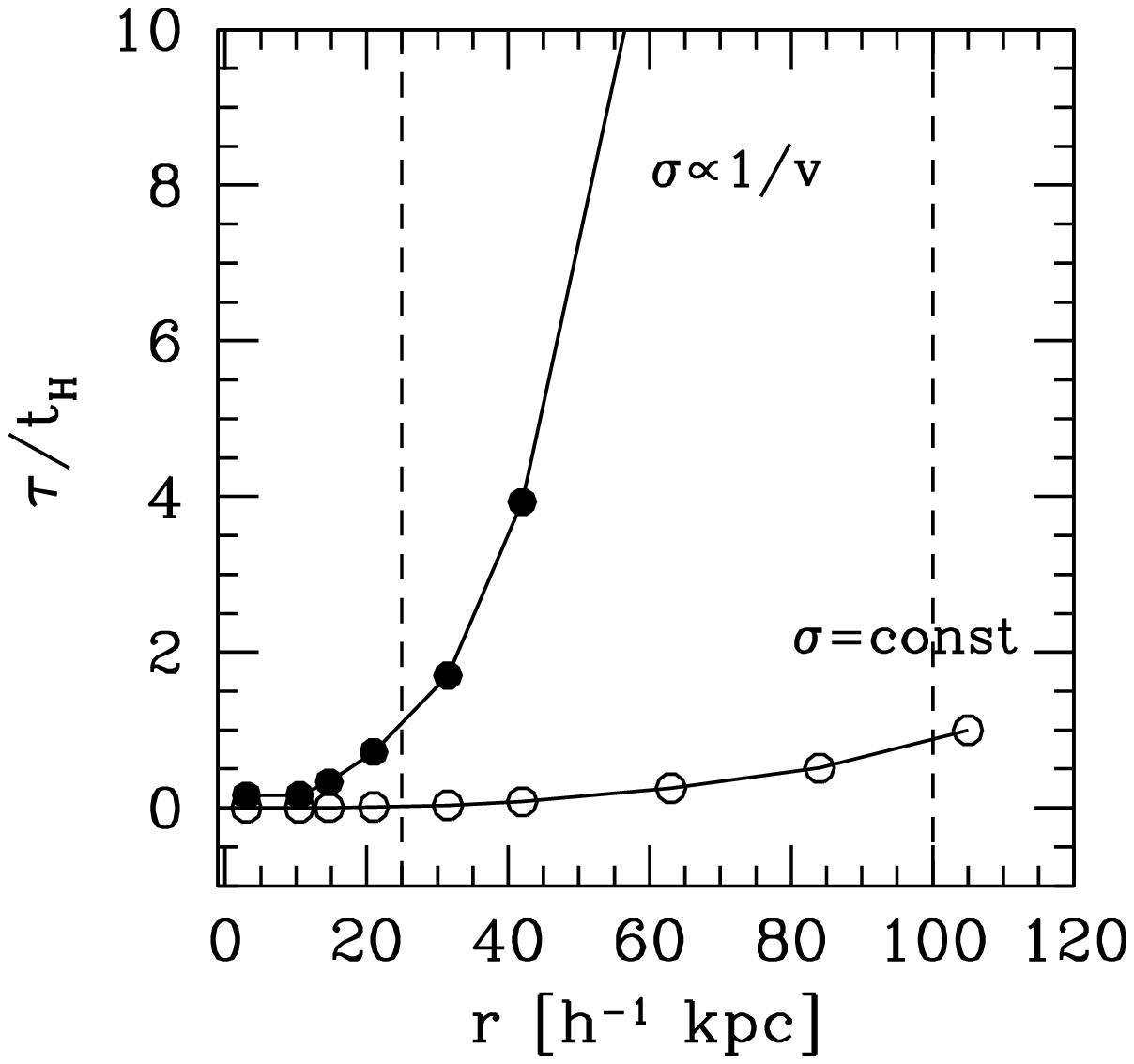}
\caption{The time required to destroy satellites in the Galactic
halo by collisions, normalized to the Hubble time, 
is shown as a function of the distance from the centre
of the Milky Way.
The filled circles show the suppression time if the cross section for self-interacting
dark matter is assumed to depend on the  halo dispersion velocity: $\sigma \propto 1/v$. 
The open circles represent the time required to destroy satellites when the cross section 
is assumed to be velocity independent: $\sigma \approx$constant.}
\end{figure}

\clearpage 

\begin{figure}
\plotone{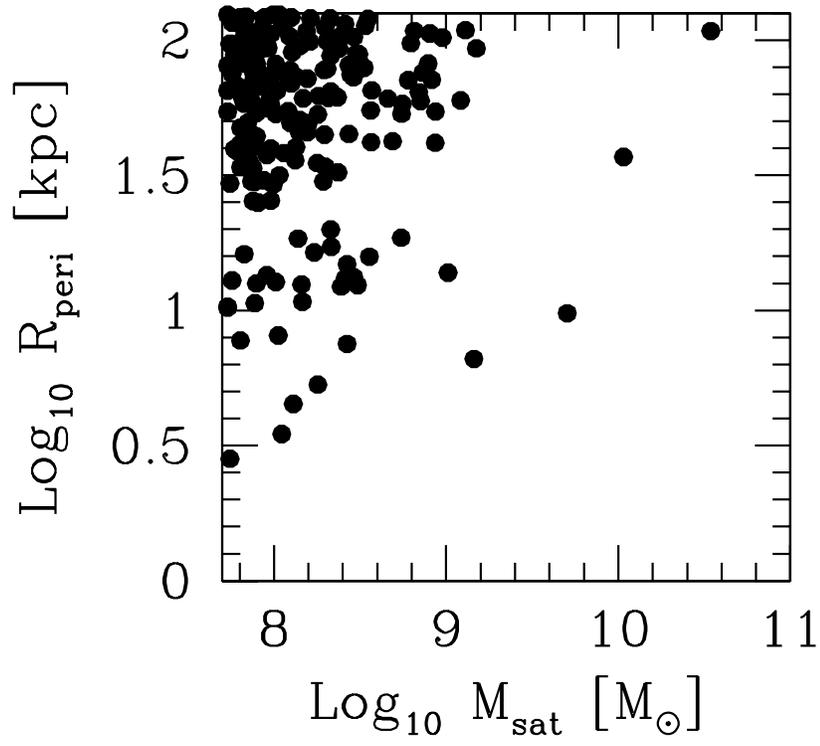}
\caption{Pericentric radii distribution vs. mass as a result of the our Monte Carlo realization
for subhalos identified in SCDM cosmological simulations.}
\end{figure}

\clearpage 

\begin{figure}
\plotone{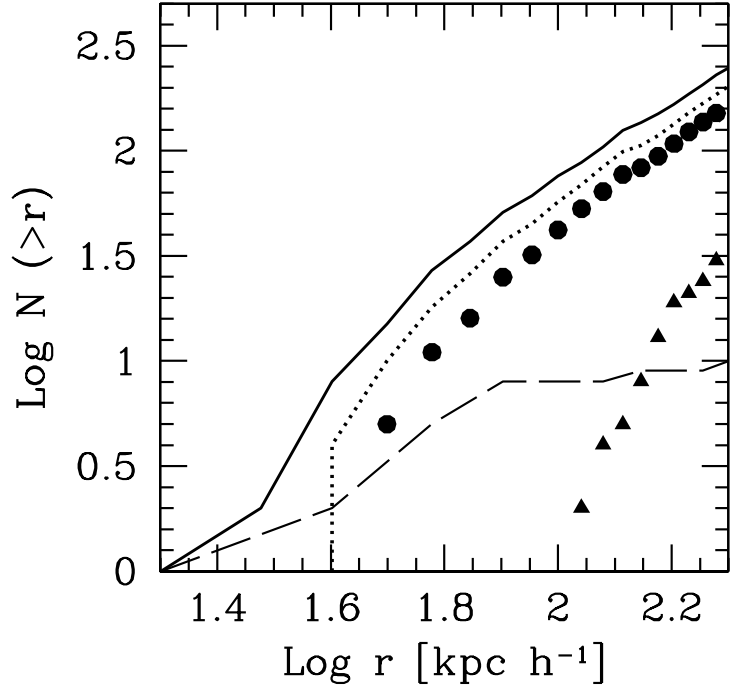}
\caption{The abundance of dark satellite halos predicted by SCDM N-body simulations 
at  different distances from the centre of the Milky Way halo are represented by the solid line
(courtesy B. Moore).
The dashed line is the cumulative number of dwarf galaxies observed
in the Local Group at different Galactocentric distances (Grebel 2000).
The dotted line is the abundance of dark satellite halos, predicted 
for a cross section dependent on the halo velocity dispersion: $\sigma \propto 1/v$,
when tidal stripping is not taken into account. 
The filled circles are the cumulative number of subhalos that survive  tidal
stripping and collisions in a self-interacting scenario with $\sigma \propto 1/v$. Triangles show that if the 
cross section is assumed to be independent of the relative velocity
the overabundance is unsolved at larger radii.}
\end{figure}

\end{document}